\documentclass[11pt]{article}
\usepackage{graphicx}
\usepackage[utf8]{inputenc} % Any characters can be typed directly from the keyboard, eg éçñ
\usepackage{textcomp} % provide lots of new symbols
\usepackage{flafter}  % Don't place floats before their definition

\usepackage{amsmath,amssymb}  % Better maths support & more symbols
\usepackage{bm}  % Define \bm{} to use bold math fonts
\usepackage{authblk}
\usepackage{memhfixc}  % remove conflict between the memoir class & hyperref
\usepackage{listings}
\numberwithin{equation}{section}
\begin{document}
\title{Exact solution of the area reactivity model of an isolated pair}
\author{Thorsten Pr\"ustel} 
\author{Martin Meier-Schellersheim} 
\affil{Laboratory of Systems Biology\\National Institute of Allergy and Infectious Diseases\\National Institutes of Health}
\maketitle
\let\oldthefootnote\thefootnote 
\renewcommand{\thefootnote}{\fnsymbol{footnote}} 
\footnotetext[1]{Email: prustelt@niaid.nih.gov, mms@niaid.nih.gov} 
\let\thefootnote\oldthefootnote 
\abstract
{
We investigate the reversible diffusion-influenced reaction of an isolated pair in two space dimensions in the context of the area reactivity model. We compute the exact Green's function in the Laplace domain for the initially unbound molecule. Furthermore, we calculate the exact expression for the Green's function in the time domain by inverting the Laplace transform via the Bromwich contour integral. The obtained results should be useful for comparing the behavior of the area reactivity model with more conventional models based on contact reactivity. 
}
\section{Introduction}
The Smoluchowski model is widely used in the theory of diffusion-influenced reactions \cite{smoluchowski:1917, Rice:1985}. According to this picture, a pair of molecules separated by a distance $r$ may react when they encounter each other at a critical distance $r=a$ via their diffusive motion. Hence, reactive molecules can be modeled by solutions of the diffusion equation that satisfy certain types of boundary conditions (BC) at the encounter distance $r=a$. In the case of an isolated pair, exact expressions for Green's functions (GF) in the time domain, describing irreversible and reversible reactions in one, two and three space dimensions, have been obtained \cite{carslaw1986conduction, Agmon:1984, TPMMS_2012JCP, kimShin:1999}. 

However, there are alternative approaches to describe the reversible diffusion-influenced reaction of an isolated pair. Ref.~\cite{Khokhlova:2012} discussed the so-called volume reactivity model that eliminates the distinct role of the encounter radius $r=a$ and instead postulates that the reaction can happen throughout the spherical volume $r\leq a$. In the present manuscript, we discuss the corresponding model in two dimensions (2D) and hence refer to it as the "area reactivity" model. 

Diffusion in 2D is special from both a conceptual and technical point of view. Conceptually, it is the critical dimension regarding recurrence and transience of random walks \cite{Toussaint_Wilczek_1983}. Technically, the mathematical
treatment appears to be more involved than in 1D and 3D \cite{TPMMS_2012JCP}.

A system of two molecules $A$ and $B$ with diffusion constants $D_{A}$ and $D_{B}$, respectively, can also be described as the diffusion of a point-like molecule with diffusion constant $D=D_{A}+D_{B}$ around a static disk. More precisely, the area-reactivity model assumes that the molecule undergoes free diffusion apart from inside the static "reaction disk" of radius $r = a$, where it may react reversibly. Without loss of generality, we assume that the disk's center is located at the origin. 
A central notion is the probability density function (PDF) $p(r,t\vert r_{0})$ that gives the probability to find the molecule unbound at a distance equal to $r$ at time $t$, given that the distance was initially $r_{0}$ at time $t=0$. Note that in contrast to the contact reactivity model, $p(r,t\vert r_{0})$ is also defined for $r<a$. Moreover, because the molecule may bind anywhere within the disk $r<a$, it makes sense to define another PDF $q(r,t\vert r_{0})$, which yields the probability to find the molecule bound at a distance equal to $r<a$ at time $t$, given that the distance was initially $r_{0}$ at time $t=0$. The rates for association and dissociation are $\kappa_{r} \Theta(a-r)p(r,t\vert r_{0})$ and $\kappa_{d}q(r,t\vert r_{0})$, respectively, where $\Theta(x)$ refers to the Heaviside step-function that vanishes for $x<0$ and assumes unity otherwise. 
Furthermore, it is assumed that the dissociated molecule is released at the same point where it assumed its bound state.

The equations of motion for the PDF  $p(r,t\vert r_{0})$ and $q(r,t\vert r_{0})$ are coupled and read \cite{Khokhlova:2012}   
\begin{eqnarray}
\frac{\partial p(r,t\vert r_{0})}{\partial t} &=& \mathcal{L}_{r}p(r , t\vert r_{0}) - \kappa_{r}\Theta(a-r)p(r,t\vert r_{0}) + \kappa_{d}q(r,t\vert r_{0}), \quad\quad\label{eqMotion1}\\
\frac{\partial q(r,t\vert r_{0})}{\partial t} &=& \kappa_{r}\Theta(a-r)p(r,t\vert r_{0}) -\kappa_{d}q(r,t\vert r_{0}), \label{eqMotion2}
\end{eqnarray}
where 
\begin{equation}
\mathcal{L}_{r} =  D\biggl(\frac{\partial^{2}}{\partial r^{2}}+\frac{1}{r}\frac{\partial}{\partial r}\biggr).
\end{equation}
The equations of motion have to be supplemented by BC at the origin and at infinity, respectively,
\begin{eqnarray}
\lim_{r\rightarrow\infty}p(r,t\vert r_{0}) &=& 0,\label{BC_Infinity}\\
\lim_{r\rightarrow 0} r \frac{\partial p(r,t\vert r_{0})}{\partial r} &=& 0 \label{BC_Zero}.
\end{eqnarray}
In the present manuscript, we focus on the case of the initially unbound molecule. Therefore, the initial conditions are
\begin{eqnarray}
2\pi r_{0} p(r, 0\vert r_{0})&=&\delta(r-r_{0}), \label{initial_bc}\\
q(r, 0\vert r_{0}) &=& 0. \label{initial_bc_q}
\end{eqnarray}
\section{Exact Green's function in the Laplace domain}
By applying the Laplace transform, Eqs.~(\ref{eqMotion1})-(\ref{eqMotion2})
become
\begin{eqnarray}
s \tilde{p}(r,s\vert r_{0}) - p(r,0\vert r_{0}) &=& \mathcal{L}_{r}\tilde{p}(r , s \vert r_{0}) - \kappa_{r}\Theta(a-r)\tilde{p}(r,s\vert r_{0})  \label{p_eqMotionLaplace}\nonumber\\
&&+ \kappa_{d}\tilde{q}(r,s\vert r_{0}),\\
s \tilde{q}(r,s\vert r_{0}) - q(r,0\vert r_{0})&=& \kappa_{r}\Theta(a-r)\tilde{p}(r,s\vert r_{0}) -\kappa_{d}\tilde{q}(r,s\vert r_{0}), \label{q_eqMotionLaplace}
\end{eqnarray}
where $s$ denotes the Laplace space variable.
We use Eq.~(\ref{initial_bc_q}) to obtain from Eq.~(\ref{q_eqMotionLaplace}) 
\begin{equation}\label{q_p_Laplace}
\tilde{q}(r,s\vert r_{0}) = \frac{\kappa_{r}}{s+\kappa_{d}}\Theta(a-r)\tilde{p}(r,s\vert r_{0}).
\end{equation}
Now we can eliminate $\tilde{q}(r,s\vert r_{0})$ from Eq.~(\ref{p_eqMotionLaplace})
\begin{equation}\label{eqMotionLaplace}
\bigg[\mathcal{L}_{r} - s - \frac{s\kappa_{r}}{s+\kappa_{d}}\Theta(a-r) \bigg]\tilde{p}(r,s\vert r_{0}) = -\frac{\delta(r-r_{0})}{2\pi r},
\end{equation}
where we used Eq.~(\ref{initial_bc}).

In the following, we will calculate the GF separately on the two different domains defined by
$r > a$ and $r<a$. The two obtained solutions will still contain unknown constants. The GF can then be completely determined by matching both expressions upon continuity requirements at $r=a$. 
Henceforth, we will denote the GF within  $r<a$ and outside $r>a$ the reactive disk by $p^{<}(r,t\vert r_{0})$ and $p^{>}(r,t\vert r_{0})$, respectively.
Also, throughout this manuscript we assume that the molecule was initially located outside the reaction area $r_{0} >  a$.

Then, we make the following ansatz for the Laplace transform of the GF $p^{>}(r,t\vert r_{0})$ outside the disk $r > a$, 
\begin{equation}\label{laplaceAnsatz}
\tilde{p}^{>}(r, s \vert r_{0}) = \tilde{p}_{0}(r, s \vert r_{0}) +\tilde{f}(r, s \vert r_{0}),
\end{equation}
where 
\begin{equation}\label{laplaceFree}
\tilde{p}_{0}(r, s \vert r_{0}) = \frac{1}{2\pi D}
\biggl\{\begin{array}{lr}
 I_{0}(vr_{0}) K_{0}(vr),&\text{$ r  >  r_{0} $} \\
 I_{0}(vr) K_{0}(qr_{0}), &\text{$r  <  r_{0}$}  
\end{array}
\end{equation}
is the Laplace transform of the free-space GF, cf. \cite[Ch.~14.8, Eq.~(2)]{carslaw1986conduction}. $I_{0}(x), K_{0}(x)$ denote the modified Bessel functions of first and second kind, respectively, and zero order \cite[Sect.~9.6]{abramowitz1964handbook}. The variable $v$ is defined by 
\begin{equation}
v:=\sqrt{s/D}.
\end{equation}
Note that the free GF takes into account the $\delta$ function term in Eq.~(\ref{eqMotionLaplace}) and therefore,
the function $\tilde{f}(r, s \vert r_{0})$ in Eq.~(\ref{laplaceAnsatz}) satisfies the Laplace transformed 2D diffusion equation \cite[Ch.~14.8, Eq.~(3)]{carslaw1986conduction}
\begin{equation}\label{laplaceDE}
\frac{d^{2}\tilde{f}}{dr^{2}} + \frac{1}{r}\frac{d\tilde{f}}{dr} - v^{2} \tilde{f} = 0.
\end{equation}
The general solution to Eq.~(\ref{laplaceDE}) is given by 
\begin{equation}
\tilde{f}(r, v) = B(s, r_{0})I_{0}(vr) + C(s, r_{0})K_{0}(vr),
\end{equation}
where $B(s, r_{0}), C(s, r_{0})$ are "constants" that may depend on $s$ and $r_{0}$.
Because we require the BC Eq.~(\ref{BC_Infinity}) and $\lim_{x\rightarrow\infty}I_{0}(x)\rightarrow \infty$, the coefficient $B(s, r_{0})$ has to vanish and the solution becomes, 
\begin{equation}\label{laplaceBR}
\tilde{f}(r, v\vert r_{0}) = C(v, r_{0}) K_{0}(vr).
\end{equation}
Next, turning to the case $r<a$, the GF satisfies
\begin{equation}\label{laplaceDE2}
\frac{d^{2}\tilde{p}^{<}}{dr^{2}} + \frac{1}{r}\frac{d\tilde{p}^{<}}{dr} - w^{2} \tilde{p}^{<} = 0,
\end{equation}
where $w$ is defined by 
\begin{equation}
w:=v\sqrt{\frac{s + \kappa_{r} + \kappa_{d}}{s+\kappa_{d}}}.
\end{equation}
Therefore, the general solution, which takes into account the BC Eq.~(\ref{BC_Zero})
is
\begin{equation}\label{generalSolution<}
p^{<}(r, w\vert r_{0}) = A(s, r_{0}) I_{0}(wr),
\end{equation}
because $\lim_{x\rightarrow 0}xK_{1}(x)\neq 0$.

The two "constants" $A(s, r_{0})$ and $C(s, r_{0})$ can be determined by the requirement that the GF and its derivative have to be continuous at $r=a$  
\begin{eqnarray}
\tilde{p}^{<}(r=a, s\vert r_{0}) &=& \tilde{p}^{>}(r=a, s\vert r_{0})\\
\frac{\partial\tilde{p}^{<}(r, s\vert r_{0})}{\partial r}\bigg\vert_{r=a} &=& \frac{\partial\tilde{p}^{>}(r, s\vert r_{0})}{\partial r}\bigg\vert_{r=a}
\end{eqnarray}
Using Eqs.~(\ref{laplaceAnsatz}),~(\ref{laplaceFree}),~(\ref{laplaceBR}), ~(\ref{generalSolution<}) as well as 
\begin{eqnarray}
I^{\prime}_{0}(x) &=& I_{1}(x), \\ 
K^{\prime}_{0}(x) &=& -K_{1}(x),\\
x^{-1}&=&I_{0}(x)K_{1}(x)+I_{1}(x)K_{0}(x),
\end{eqnarray}
\cite[Eqs.~(9.6.27),~(9.6.15)]{abramowitz1964handbook}, we obtain
\begin{eqnarray}
A(s,r_{0}) &=& \frac{K_{0}(vr_{0})}{2\pi a D \mathcal{N}}, \\
C(s,r_{0}) &=& \frac{K_{0}(vr_{0})}{2\pi a D K_{0}(va)}\bigg[ \frac{I_{0}(wa)}{\mathcal{N}} - aI_{0}(va)\bigg], 
\end{eqnarray}
where we introduced
\begin{eqnarray}
\mathcal{N} = v I_{0}(wa)K_{1}(va) + w I_{1}(wa)K_{0}(va).
\end{eqnarray}
\section{Exact Green's function in the time domain}
\begin{figure}
\includegraphics[scale=0.45]{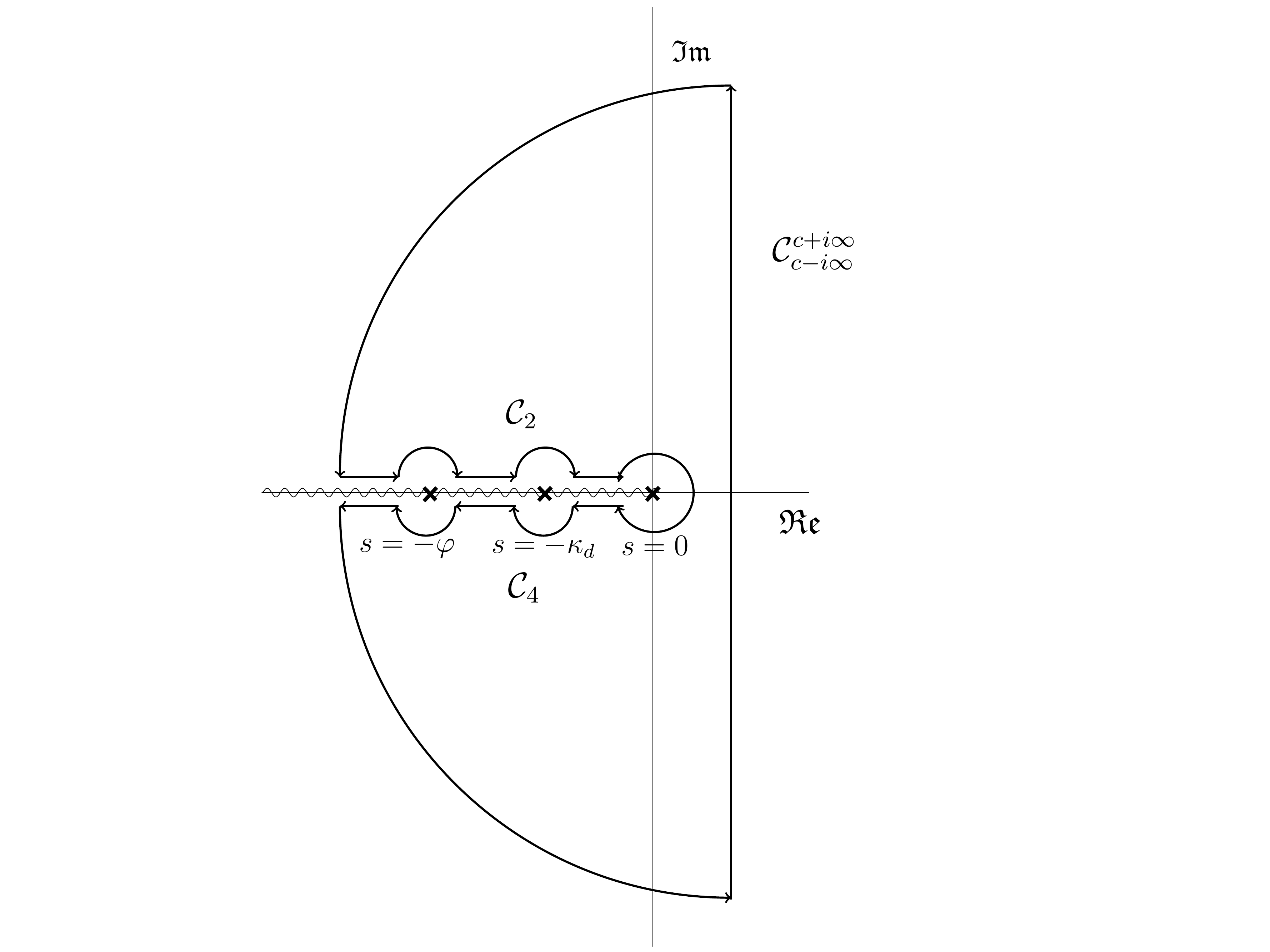}
\caption{Integration contour used for 
calculating the GF 
in the time domain, Eq.~(\ref{cauchy}).}\label{fig:contour}
\end{figure}

To find the corresponding expressions for $p^{<}(r,t\vert r_{0}), p^{>}(r,t\vert r_{0})$ in the time domain, we apply the inversion theorem for the Laplace transformation
\begin{equation}\label{inversionFormula}
p^{<}(r, t \vert r_{0}) = \frac{1}{2\pi i} \int^{c+i\infty}_{c-i\infty} e^{st}\,\tilde{p}^{<}(r, s\vert r_{0} )ds.
\end{equation}
We note that $\tilde{p}^{<}(r, s\vert r_{0} )$ has three branch points at $s=0, -\kappa_{d}$ and $s = -\kappa_{r}-\kappa_{d} \equiv -\varphi$. Therefore, to calculate the Bromwich integral, we use the contour of Fig.~\ref{fig:contour} with a branch cut along the negative real axis, cf. \cite[Ch. 12.3, FIG. 40]{carslaw1986conduction}. We arrive at 
\begin{eqnarray}\label{cauchy}
\int^{c+i\infty}_{c-i\infty} e^{st}\,\tilde{p}^{<}(r, s\vert r_{0} )ds 
=&-&\int_{\mathcal{C}_{2}} e^{ps}\,\tilde{p}^{<}(r, s\vert r_{0} )ds  \nonumber\\
&-& \int_{\mathcal{C}_{4}} e^{st}\,\tilde{p}^{<}(r, s\vert r_{0} )ds.
\end{eqnarray}
To calculate the integral $\int_{\mathcal{C}_{2}}$, we choose
$
s = D x^{2} e^{i \pi }.
$
Then,
\begin{eqnarray}
v &=& ix, \quad\text{for}\,\, s\in ]-\infty, 0[\\
w &=& ix\sqrt{\frac{ Dx^{2} - \varphi}{Dx^{2} - \kappa_{d} }}\equiv ix\xi_{1} \quad\text{for}\,\, s\in ]-\infty, -\varphi[,\\
w &=& x\sqrt{\frac{\varphi - Dx^{2}}{Dx^{2} - \kappa_{d}}} \equiv x\xi_{2} \quad\text{for}\,\, s\in]-\varphi, -\kappa_{d}[,\\
w &=& ix\sqrt{\frac{\varphi - Dx^{2}}{\kappa_{d} - Dx^{2}}}= ix\xi_{1} \quad\text{for}\,\, s\in ]-\kappa_{d}, 0[,\\
\end{eqnarray}
We now make use of \cite[Append.~3, Eqs.~(25), (26))]{carslaw1986conduction}
\begin{eqnarray}
I_{n}(xe^{\pm \pi i/2}) &=& e^{\pm n\pi i/2} J_{n}(x), \\
K_{n}(xe^{\pm \pi i/2}) &=& \pm\frac{1}{2}\pi i e^{\mp n\pi i/2} [-J_{n}(x) \pm i Y_{n}(x)].
\end{eqnarray}
$J_{n}(x), Y_{n}(x)$ refer to the Bessel functions of first and second kind, respectively \cite[Sect.~9.1]{abramowitz1964handbook}.
It follows that
\begin{eqnarray}
&&\int_{\mathcal{C}_{2}}e^{st}\,\tilde{p}^{<}(r, s\vert r_{0} )ds  =   \frac{1}{\pi a}\bigg[\int^{\sqrt{\frac{\varphi}{D}}}_{\sqrt{\frac{\kappa_{d}}{D}}}e^{-Dx^{2}t}g^{(2)}(r,r_{0}, x) dx \nonumber\\
&& - \int^{\sqrt{\frac{\kappa_{d}}{D}}}_{0}e^{-Dx^{2}t}g^{(1)}(r,r_{0}, x) dx - \int^{\infty}_{\sqrt{\frac{\varphi}{D}}}e^{-Dx^{2}t}g^{(1)}(r,r_{0}, x) dx\bigg],\quad\quad\quad
%&&-\frac{1}{\pi a}\int^{\infty}_{0}e^{-Dx^{2}t}[Y_{0}(xr_{0})+ i J_{0}(xr_{0})]J_{0}(\xi xr) \int^{\infty}_{0}e^{-Dx^{2}t}[Y_{0}(xr_{0})+ i J_{0}(xr_{0})]J_{0}(\xi xr) \nonumber \\
%&& \times &\frac{[\rho(x)-i\psi(x)]}{\rho(x)^{2} + \psi(x)^{2}}   dx,
\end{eqnarray}
where we introduced 
\begin{eqnarray}
 g^{(2)}(r, r_{0}, x) &\equiv& g^{(2)}_{R}(r, r_{0}, x) + i  g^{(2)}_{I}(r, r_{0}, x), \nonumber \\
& = & I_{0}(x\xi_{2}r)\frac{\eta(r_{0}) +  i \lambda(r_{0})}{\alpha^{2} + \beta^{2}},\quad\quad \\
 g^{(1)}(r, r_{0}, x) &\equiv& g^{(1)}_{R}(r, r_{0}, x) + i  g^{(1)}_{I}(r, r_{0}, x), \nonumber \\
& = & J_{0}(x\xi_{1}r)\frac{\omega(r_{0}) +  i \varkappa(r_{0})}{\gamma^{2} + \delta^{2}},\quad\quad 
\end{eqnarray}
and
\begin{eqnarray}
\eta(r_{0}) &=& \alpha Y_{0}(xr_{0})+ \beta J_{0}(xr_{0}),\\
\lambda(r_{0}) &=& \alpha J_{0}(xr_{0}) - \beta Y_{0}(xr_{0}),\\
\alpha &=& \xi_{2} I_{1}(\xi_{2} xa)Y_{0}(xa) + I_{0}(\xi_{2} xa)Y_{1}(xa),\\
\beta &=& \xi_{2} I_{1}(\xi_{2} xa)J_{0}(xa) + I_{0}(\xi_{2} xa)J_{1}(xa),\\
\omega(r_{0}) &=& \gamma Y_{0}(xr_{0})+ \delta J_{0}(xr_{0}),\\
\varkappa(r_{0}) &=&  \gamma J_{0}(xr_{0}) -\delta Y_{0}(xr_{0}),\\
\gamma &=& \xi_{1} J_{1}(\xi_{1} xa)Y_{0}(xa) - J_{0}(\xi_{1} xa)Y_{1}(xa), \\
\delta &=& \xi_{1} J_{1}(\xi_{1} xa)J_{0}(xa) - J_{0}(\xi_{1} xa)J_{1}(xa). 
\end{eqnarray}
Now, to calculate the integral along the contour $\mathcal{C}_{4}$, we choose $s = Dx^{2}e^{-i\pi}$ and after an analogous calculation one finds that
\begin{equation}
\int_{\mathcal{C}_{2}}e^{st}\,\tilde{p}^{<}(r, s\vert r_{0} )ds=-\bigg(\int_{\mathcal{C}_{4}}e^{st}\,\tilde{p}^{<}(r, s\vert r_{0} )ds\bigg)^{\ast},
\end{equation}
where $\ast$ denotes complex conjugation.
Thus, one obtains for the GF $p^{<}(r, t \vert r_{0})$ on the domain $r<a$
\begin{eqnarray}\label{GF<}
p^{<}(r, t \vert r_{0}) &=& -\frac{1}{\pi} \mathfrak{Im}\bigg(\int_{\mathcal{C}_{2}}e^{st}\,\tilde{p}^{<}(r, s\vert r_{0} )ds\bigg)\nonumber \\
&=&-\frac{1}{\pi^{2} a}\bigg[\int^{\sqrt{\frac{\varphi}{D}}}_{\sqrt{\frac{\kappa_{d}}{D}}}e^{-Dx^{2}t}g^{(2)}_{I}(r,r_{0}, x) dx \nonumber\\
&-& \int^{\sqrt{\frac{\kappa_{d}}{D}}}_{0}e^{-Dx^{2}t}g^{(1)}_{I}(r,r_{0}, x) dx - \int^{\infty}_{\sqrt{\frac{\varphi}{D}}}e^{-Dx^{2}t}g^{(1)}_{I}(r,r_{0}, x) dx\bigg],\quad\quad\quad
\end{eqnarray}
Analogously, we can proceed to compute the GF for the region $r>a$.
Therefore, we only give the result
\begin{eqnarray}\label{GF>}
&&p^{>}(r, t \vert r_{0}) = \frac{1}{4\pi D t}
 e^{-(r^{2}+r^{2}_{0})/4Dt}I_{0}\bigg(\frac{rr_{0}}{2Dt}\bigg)\nonumber \\
&& +\frac{1}{\pi^{2}a}\bigg[\int^{\sqrt{\frac{\kappa_{d}}{D}}}_{0} e^{-Dx^{2}t} h^{(1)}(r,r_{0}, x) dx + \int^{\infty}_{\sqrt{\frac{\varphi}{D}}} e^{-Dx^{2}t} h^{(1)}(r,r_{0}, x) dx\quad\quad\quad\nonumber\\
&& -\int^{\sqrt{\frac{\varphi}{D}}}_{\sqrt{\frac{\kappa_{d}}{D}}}e^{-Dx^{2}t} h^{(2)}(r,r_{0}, x) dx\bigg] -\frac{1}{2\pi}\int^{\infty}_{0}e^{-Dx^{2}t} h^{(3)}(r,r_{0}, x)xdx,
%&&+ \frac{1}{\pi^{2}a}\int^{\infty}_{0}e^{-Dx^{2}t} J_{0}(\xi x a) \frac{\rho \Lambda - \psi \Xi}{[\rho^{2} + \psi^{2}][J^{2}_{0}(xa) + Y^{2}_{0}(xa)]}   dx,\qquad
\end{eqnarray}
where we defined
\begin{eqnarray}
h^{(1)}(r,r_{0}, x) &=& J_{0}(x\xi_{1} a)\frac{\rho(r) \omega(r_{0})+ \psi(r) \varkappa(r_{0})}{[\gamma^{2}+\delta^{2}] [J_{0}^{2}(xa)+Y_{0}^{2}(xa)]} ,\\
h^{(2)}(r,r_{0}, x) &=& I_{0}(x\xi_{2} a)\frac{\rho(r) \eta(r_{0})+ \psi(r) \lambda(r_{0})}{[\alpha^{2}+\beta^{2}] [J_{0}^{2}(xa)+Y_{0}^{2}(xa)]},\\
h^{(3)}(r,r_{0}, x) &=& J_{0}(xa)\frac{\Pi(r, r_{0}) Y_{0}(xa)+ \Omega(r, r_{0}) J_{0}(xa)}{J_{0}^{2}(xa)+Y_{0}^{2}(xa)},
\end{eqnarray}
and
\begin{eqnarray}
\rho(r) &=& J_{0}(xr)Y_{0}(xa) - Y_{0}(xr)J_{0}(xa),\\
\psi(r) &=& J_{0}(xr)J_{0}(xa) + Y_{0}(xr)Y_{0}(xa), \\
\Omega(r, r_{0}) &=& J_{0}(xr)J_{0}(xr_{0}) - Y_{0}(xr)Y_{0}(xr_{0}), \\
\Pi(r, r_{0})  &=& Y_{0}(xr)J_{0}(xr_{0}) + J_{0}(xr)Y_{0}(xr_{0}).
\end{eqnarray}
Note that the first term appearing on the rhs of Eq.~(\ref{GF>}) is the inverse Laplace transform of Eq.~(\ref{laplaceFree}), cf.~\cite[Ch.~14.8, Eq.~(2)]{carslaw1986conduction}.

Finally, we can compute an exact expression for $q(r,t\vert r_{0})$ by virtue of  Eq.~(\ref{q_p_Laplace}) and the convolution theorem of the Laplace transform.
We obtain for $r < a$
\begin{eqnarray}
q(r, t \vert r_{0}) &=& -\frac{1}{\pi^{2} a}\bigg[\int^{\sqrt{\frac{\varphi}{D}}}_{\sqrt{\frac{\kappa_{d}}{D}}}\bigg(\frac{e^{-Dx^{2}t}-e^{-\kappa_{d}x^{2}t}}{\kappa_{d}-Dx^{2}}\bigg)g^{(2)}_{I}(r,r_{0}, x) dx \nonumber\\
&-& \int^{\sqrt{\frac{\kappa_{d}}{D}}}_{0}\bigg(\frac{e^{-Dx^{2}t}-e^{-\kappa_{d}t}}{\kappa_{d}-Dx^{2}}\bigg)g^{(1)}_{I}(r,r_{0}, x) dx \nonumber\\
&-& \int^{\infty}_{\sqrt{\frac{\varphi}{D}}}\bigg(\frac{e^{-Dx^{2}t}-e^{-\kappa_{d}t}}{\kappa_{d}-Dx^{2}}\bigg)g^{(1)}_{I}(r,r_{0}, x) dx\bigg]. \quad\quad\quad
\end{eqnarray}
Clearly, $q(r, t \vert r_{0})$ vanishes for $r > a$.
 
The case of an initially unbound molecule with $r_{0} < a$ and the case of the initially bound molecule will be considered in a forthcoming manuscript.
\\
%\newpage
\section*{Acknowledgments}
This research was supported by the Intramural Research Program of the NIH, National Institute of Allergy and Infectious Diseases.

%\bibliographystyle{plain} 
%\bibliography{EG}

%\begin{thebibliography}
%\end{thebibliography}%
\end{document}